\documentclass[nofootinbib,prl,superscriptaddress,preprint,tightenlines,preprintnumbers]{revtex4}
\usepackage{graphicx}
\usepackage{amsfonts}
\usepackage{amssymb}
\usepackage{amsmath}

\newcommand{\be}{\begin{equation}}
\newcommand{\ee}{\end{equation}}
\newcommand{\ba}{\begin{eqnarray}}
\newcommand{\ea}{\end{eqnarray}}

\newcommand*{\di}{\partial}
\newcommand{\SNR}{\text{SNR}}

\begin{document}

\preprint{astro-ph/0610470}

\title{A gravitational wave window on extra dimensions}

\author{Chris Clarkson}
\email{chris.clarkson@port.ac.uk}

\affiliation{Cosmology and Gravity Group, Department of Mathematics
and Applied Mathematics, University of Cape Town, Rondebosch 7701,
Cape Town, South Africa}

\affiliation{Institute of Cosmology \& Gravitation, University of
Portsmouth, Portsmouth~PO1~2EG, UK}

\author{Sanjeev S.~Seahra}
\email{sanjeev.seahra@port.ac.uk}

\affiliation{Institute of Cosmology \& Gravitation, University of
Portsmouth, Portsmouth~PO1~2EG, UK}

\setlength\arraycolsep{2pt}

\begin{abstract}


We report on the possibility of detecting a submillimetre-sized
extra dimension by observing gravitational waves (GWs) emitted by
pointlike objects orbiting a braneworld black hole. Matter in the
`visible' universe can generate a discrete spectrum of high
frequency GWs with amplitudes moderately weaker than the
predictions of general relativity (GR), while GW signals generated
by matter on a `shadow' brane hidden in the bulk are potentially
strong enough to be detected using current technology. We know of
no other astrophysical phenomena that produces GWs with a similar
spectrum, which stresses the need to develop detectors capable of
measuring this high-frequency signature of large extra dimensions.

\end{abstract}


\date{16 October, 2006}

\maketitle


\paragraph*{\bf Motivation}

String theory-inspired braneworld models
\cite{Horava:1996ma,Randall:1999ee} envisage our universe as a 4D
membrane embedded in some higher-dimensional space. Standard Model
particles and fields are assumed to be confined to the `brane',
while gravity propagates in the higher-dimensional `bulk'. The
principal observational features of such models take the form of
modifications to 4D gravity. For static situations, we expect
deviations from Newton's law at distances less than the curvature
scale of the bulk, $\ell$. Precision laboratory measurements yield
that $\ell \lesssim 0.1\,\text{mm}$ \cite{Adelberger:2003zx}.
Similarly, as we demonstrate here, in the case of dynamic
gravitational fields we can expect significant higher-dimensional
effects at \emph{frequencies} in excess of $\sim c/\ell$. Presently,
there is a vigorous worldwide effort to build detectors capable of
observing dynamic gravitational degrees of freedom
(i.e.,~gravitational waves) and hence verify one of the last
untested predictions of Einstein's general relativity. An intriguing
question is how the braneworld paradigm may affect what these
detectors see, and whether or not we can obtain useful constraints
on extra dimensions.  To address these issues, we need to concretely
model how GWs are generated in braneworld scenarios. The purpose of
this work is to predict the spectrum and amplitude of GWs generated
by pointlike bodies orbiting a braneworld black hole.

\paragraph*{\bf The black string braneworld}

As in our previous work with R.~Maartens~\cite{2005PhRvL..94l1302S},
we model a braneworld black hole as a 5D black string spacetime
between a `visible' brane at $y = 0$ and a `shadow' brane at $y=d$.
The metric is given by $ ds_5^2 = e^{-2|y|/\ell} ds_{\text{Schw}}^2
+ dy^2,$ where $ds_{\text{Schw}}^2$ is the 4D Schwarzschild line
element.  We require $d/\ell\gtrsim5$ to comply with post-Newtonian
solar system constraints~\cite{Garriga:1999yh}. This looks like the
Schwarzschild solution on `our' visible brane, with deviations from
GR appearing perturbatively. This solution is stable if the mass of
the string $M$ is large compared to the brane separation:
${M}/{M_\odot} \gtrsim 1.1 \times 10^{-6} ( {\ell}/ {0.1\,\text{mm}}
) \, e^{(d-5\ell)/\ell}$.  If the string is too light an instability
shows up in the spherical perturbations, but not for higher
multipoles~\cite{1993PhRvL..70.2837G,2006hep.th....2001K}.

\paragraph*{\bf Kaluza-Klein radiation from orbiting
particles}

A small compact object (which we model as a delta function) of mass
$M_p$ on either brane will orbit the black string in the same way as
it would a Schwarzschild black hole in 4 dimensions. As in GR, it
will emit massless spin-2 gravitational radiation, but unlike GR it
will also generate fluctuations in the brane's position as well
`Kaluza-Klein' (KK) modes.  These are 5D massless GWs that have
momentum along the extra dimension, and so behave like a coupled
system of spin-0, spin-1, and spin-2 \emph{massive} fields on either
brane. Owing to the finite separation of the branes, the spectrum of
KK masses $m_n$ is discrete. We concentrate on the spherical
component of KK radiation generated by particles orbiting the string
on either the visible or shadow brane.

\begin{figure}
\includegraphics[width=\columnwidth]{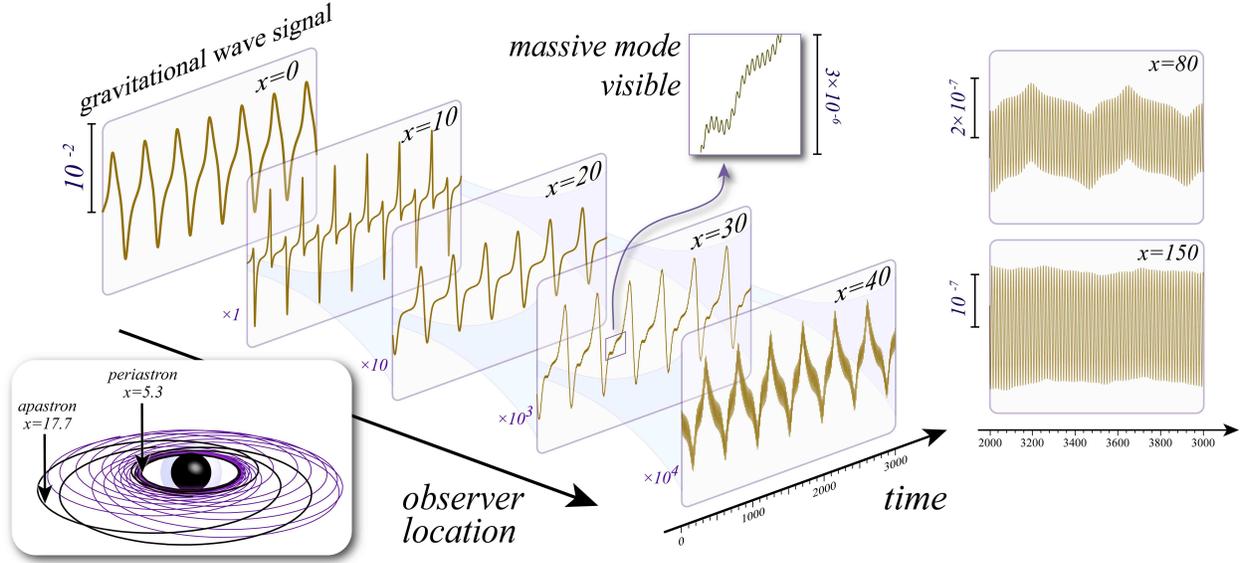}
\caption{\textbf{Radiation from a periodic orbit.} The steady-state
KK gravitational wave signal induced by a particle undergoing a
periodic orbit around the black string with $\mu=0.5$. The orbit
(\emph{bottom left}) has eccentricity 0.5 and angular momentum 3.62
in the notation of
Refs.~\cite{2004PhRvD..69d4025M,PhysRevD.50.3816}. The waveform of
radiation falling into the black string is quite different than that
of radiation escaping to infinity:  The infalling signal precisely
mimics the orbital profile of the source, while the outgoing signal
is dominated by monochromatic radiation whose frequency is
proportional to the KK mass $f = m c^2/2\pi\hbar$. 
\label{fig:flower.orbit.figures.all.tiny.ps}}
\end{figure}

\begin{figure}
\includegraphics[width=\columnwidth]{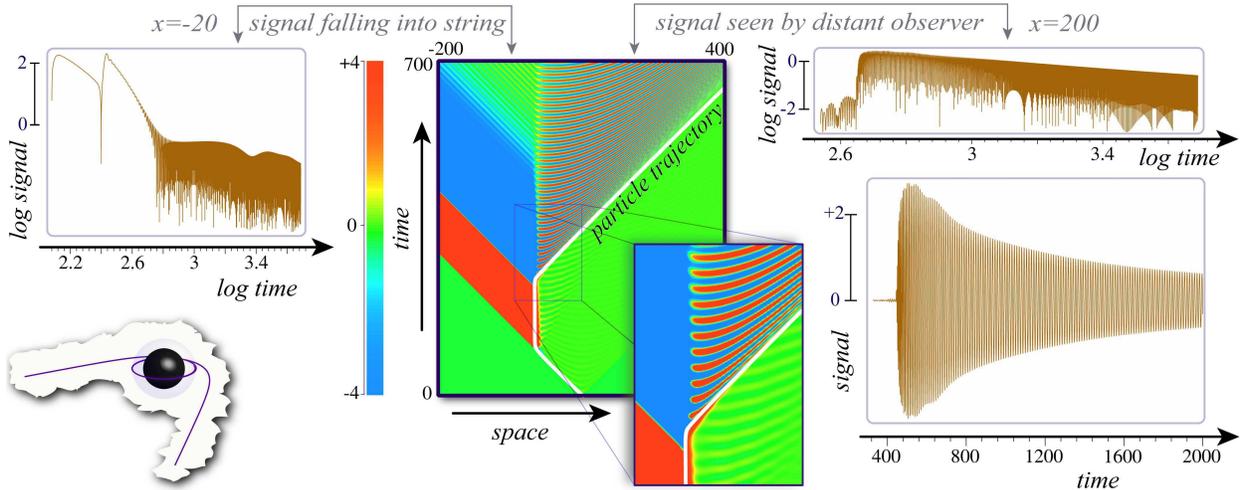}
\caption{\textbf{Radiation from a fly-by orbit} The massive mode
($\mu = 0.5$) GW signal from a fly-by orbit (\emph{bottom left})
with eccentricity $3.0$ and angular momentum $\sim 3700$.  We show
the generated radiation in a spacetime diagram (\emph{center}), as
seen by an observer close to the string (\emph{top left}), and as
seen by a distant observer (\emph{right}).
\label{fig:fly-by.orbit...figures.2.ps}}
\end{figure}

As described in the Appendix, spherical KK modes are governed by a
pair of one-dimensional coupled wave equations sourced by the
particle.  To illustrate typical waveforms seen by distant
observers, we numerically integrate these equations in the case when
the dimensionless KK mass is $\mu = GMm/\hbar c = 0.5$.  We consider
two types of source trajectory: an eccentric periodic orbit
(Fig.~\ref{fig:flower.orbit.figures.all.tiny.ps}), and a `fly-by'
orbit (Fig.~\ref{fig:fly-by.orbit...figures.2.ps}). In the former
instance a nearly monochromatic steady-state signal is seen far from
the string, while in the latter case we see a burst of radiation
followed by a slowly decaying tail.

\begin{figure}
\includegraphics[width=0.6\textwidth]{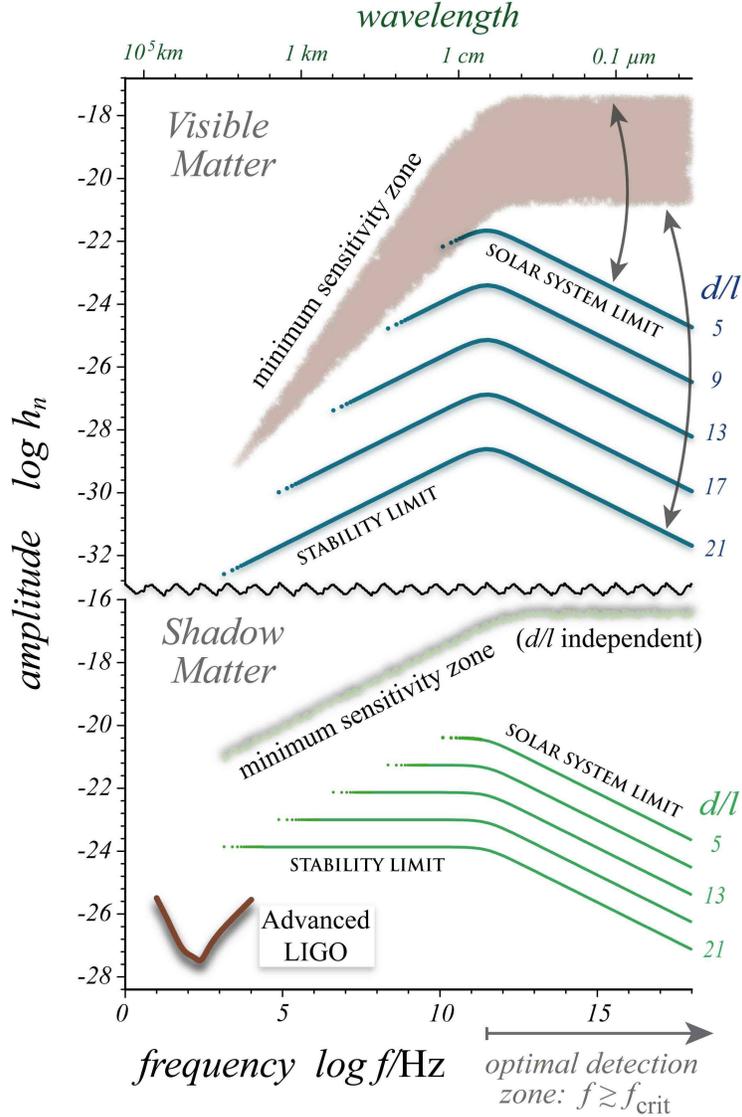}
\caption{\textbf{Massive mode amplitudes for visible and shadow
sources.}  We have taken $\ell=0.1\,$mm, the string mass to be
$10\,M_\odot$, the particle mass to be $1.4\,M_\odot$, source
distance to be $1\,$kpc, and $\mathcal{A} = 1$. For comparison, the
dimensionless one-year sensitivity curve for Advanced LIGO
(ALIGO)~\cite{LIGO} is shown~\cite{2005PhRvD..71h4008A}. The shaded
region shows the minimum $h_\text{strain}\,\text{Hz}^{1/2}$ required
to detect the KK signal with $\SNR \ge 5$ as a function of detector
frequency $f_0$, assuming a one-year integration time and a periodic
source. \label{fig:amplitudes.small.ps}}
\end{figure}

For a given source, it is possible to estimate the amplitude $h_n$
of each massive mode as measured on earth.  We show these
amplitudes as a function of their frequency $f_n$ in
Fig.~\ref{fig:amplitudes.small.ps} for a number of different
cases.  Note that the KK frequencies are bounded from below:
$
    f_n \ge f_\text{min} \sim 12 \,\text{GHz} \left( {\ell}/{0.1\,\text{mm}}
    \right)^{-1} e^{-(d - 5\ell)/\ell}
$
. The variation of amplitude with frequency is qualitatively
different for $f_n$ greater or less than a critical value:
$
    f_\text{crit} = {c}/{\pi^2\ell} \sim 304\,\text{GHz} \left( {\ell}/{0.1\,\text{mm}}
    \right)^{-1}
$
. For `visible' sources located on our brane, the amplitudes are
peaked for $f_n \sim f_\text{crit}$.  For `shadow' sources located
on the other brane, the $h_n$ are both independent of frequency and
maximised for $f_n \lesssim f_\text{crit}$.  In either case, the
amplitudes are bounded from above $h_n \lesssim h_\text{max}$, where
\begin{multline}\label{amplitudes}
    h_\text{max} \sim \mathcal{A} \left( \frac{M_p}{M_\odot} \right)
    \left( \frac{r}{\text{kpc}} \right)^{-1}
    \left( \frac{M}{M_\odot} \right)^{-1/2}  \left(
    \frac{\ell}{0.1\,\text{mm}} \right)^{1/2} \\ \times
    \begin{cases}
        5.0\times 10^{-22} e^{-(d-5\ell)/\ell}, & \text{visible source},\\
        9.1\times 10^{-21} e^{-(d-5\ell)/2\ell}, & \text{shadow source}.
    \end{cases}
\end{multline}
Here, $r$ is the distance to the string and $\mathcal{A}$ is the
characteristic signal amplitude determined from simulations.  We
find that $\mathcal{A} \sim 1$ for fly-by orbits and $\mathcal{A}
\sim 10^{-6}$ for periodic orbits.  In the former case,
$h_\text{max}$ is comparable to the expected signal strength in
GR.

\paragraph*{\bf Detection scenarios}

Consider a GW detector whose sensitivity is characterised by the
spectral noise density $S(f)$.  If $H_n(t)$ is the detector's
(linear) response to the $n^\text{th}$ KK mode, then the
signal-to-noise ratio for the \emph{total} massive mode signal
built-up over an observation time $T$
is $
    \SNR = [ \sum_n {2}{S^{-1}(f_n)} \int_0^{T} H_n^2(t)\, dt ]^{1/2}
 $
~\cite{2000PhRvD..61f2001J}. Since the frequency separation $\sim
f_\text{crit} e^{-d/\ell}$ between KK modes is small in most cases,
this sum can approximated by an integral.  This leads to the
semi-empirical formula:
\begin{equation}\label{SNR}
    \SNR \approx \Gamma e^{d/2\ell} h_\text{max}
    \begin{cases}
        0.45 \, T^{1/2}, & \text{periodic orbit},\!\!\! \\
        0.83 \, T^{5/6}_\text{str} f_\text{crit}^{1/3},
        & \text{fly-by orbit},
    \end{cases}
\end{equation}
where $T_\text{str} = 2GM/c^3 \approx 4.9
(M/M_\odot)\,\mu\text{s}$ is the characteristic timescale set by
the string mass, which is typically much less than the observation
time. The pre-factor depends on the detector, particle orbit, and
source location:
\begin{equation}\label{Gamma}
    \Gamma = \Gamma[S,d,\ell] = \left[ \int_{32.4 e^{-d/\ell}}^\infty
    \frac{Q(u)}{S(f_\text{crit}{u})} du \right]^{1/2},
\end{equation}
where
\begin{equation}
    \begin{array}{ccc}
        \hline Q(u) & \text{periodic orbit} & \text{fly-by orbit} \\
        \hline \hline \text{visible source} & 2u(1+u^2)^{-1} & 2u^{5/3}(1+u^2)^{-1}  \\
        \text{shadow source} & (1+u^2)^{-1/2} & u^{2/3} (1+u^2)^{-1/2} \\
        \hline
    \end{array}
\end{equation}
Detectors which maximise $\Gamma$ stand the best chance of
detecting the KK signal.

\begin{figure}
\includegraphics[width=0.5\textwidth]{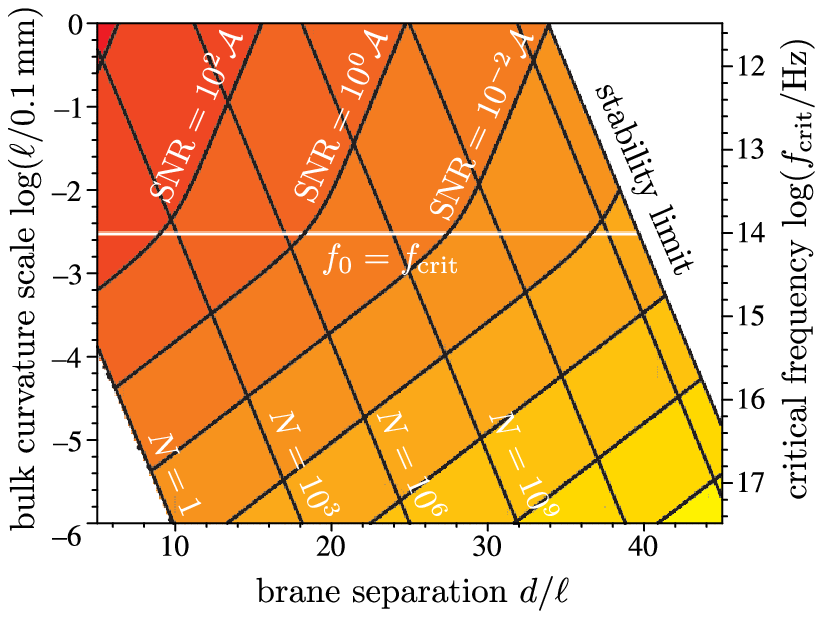}
\caption{\textbf{Observations in the extra-dimensional parameter space.}
Signal-to-noise prediction for massive mode GWs generated by a
fly-by orbit on the visible brane for a high-frequency detector
with $f_0 = 10^{14}\,\text{Hz}$, $\Delta f = 10^{13}\,\text{Hz}$,
and $h_\text{strain} = 10^{-23}\,\text{Hz}^{-1/2}$.  The particle
mass is $1\,M_\odot$ while the string corresponds to the `black
hole' at the galactic centre ($M\approx4 \times 10^6\,M_\odot$ and
$r \approx 8\,\text{kpc}$); the scaling of $\SNR$ with $M_p$ and
$r$ is given by (\ref{amplitudes}). Also shown are the number of
KK modes $N$ in this detector's range as a function of $\ell$ and
$d/\ell$. Assuming that the other parameters can be obtained from
other observations (e.g.~the zero-mode GW signal), simultaneous
measurements of $\SNR$ and $N$ can determine the bulk curvature
scale and brane separation uniquely. \label{fig:SNR visible}}
\end{figure}

The simplest noise model for a GW detector is one in which the
characteristic strain sensitivity $h_\text{strain} = \sqrt{S(f)}$
is constant over a band $[f_0 - \Delta f,f_0 + \Delta f]$, and is
otherwise infinite.  In Fig.~\ref{fig:amplitudes.small.ps}, we
show the $h_\text{strain}$ required of such a detector in order to
observe periodic-orbit KK radiation as a function of $f_0$. We
hold the logarithmic bandwidth of the detector constant, which
yields that the minimum sensitivity is actually independent of
$f_0$ for $f_0 \gtrsim f_\text{crit}$.  That is, the optimal means
of detecting this type of KK radiation is via a high-frequency GW
detector with $f_0 \gtrsim f_\text{crit}$.  Several designs for
devices approaching this zone have been proposed or implemented
\cite{Cruise:2000za,Ballantini:2003nt}. Fig.~\ref{fig:SNR visible}
shows how such a detector can be used to constrain the fundamental
parameters of our model.

\begin{figure}
\includegraphics[width=0.5\textwidth]{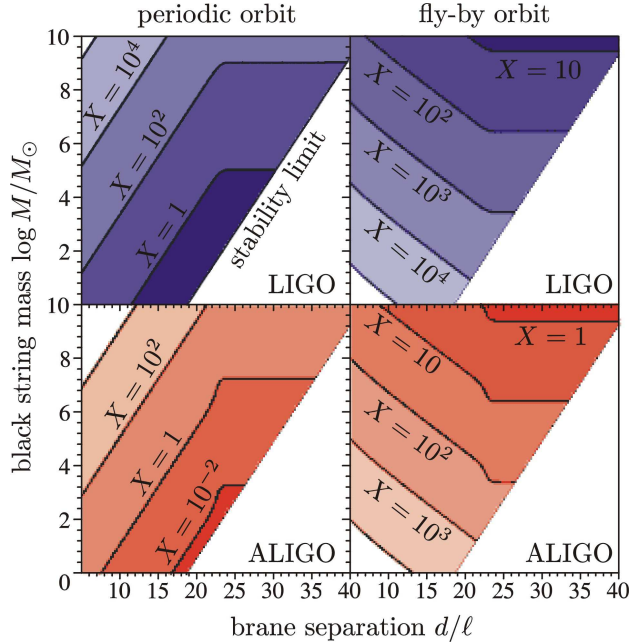}
\caption{\textbf{Prospects for detecting KK radiation from shadow matter
particles by LIGO and ALIGO.} For any black string-shadow particle
interaction, we can assign a detectability statistic $X =
\mathcal{A}(M_p/M_\odot)(r/\text{kpc})^{-1}$. The contours
indicate events that are detectable with $\SNR \ge 1$ (assuming a
one-year integration time in the case of periodic sources, and
$\ell = 0.1\,$mm).  We have assumed that the detector sensitivity
is entirely photon shot-noise/radiation pressure limited for $f
\gtrsim 10^4\,\text{Hz}$. \label{fig:SNR shadow}}
\end{figure}

Eqs.~(\ref{amplitudes}) and (\ref{SNR}) imply that the
detector-independent ratio $\SNR/\Gamma$ decreases exponentially
with $d/\ell$ for visible matter, but is independent of brane
separation for shadow matter. Furthermore, as seen in
Fig.~\ref{fig:amplitudes.small.ps}, the KK amplitudes $h_n$
generated by shadow matter are not suppressed for $f_n \lesssim
f_\text{crit}$, unlike the visible matter case.  These facts imply
that a direct detection of shadow matter is within the capability
of `low-frequency' devices such as LIGO and ALIGO, which are
otherwise insensitive to KK radiation from realistic sources on
our brane. In Fig.~\ref{fig:SNR shadow}, we show the types of
events detectable by these two interferometers when they achieve
their respective design sensitivities.

\paragraph*{\bf Discussion}

Gravitational waves may well prove to be a critical tool in
placing limits on, or detecting positive signatures of, large
extra dimensions. By using a simple model of a braneworld black
hole, we have shown how brane-confined matter can generate GW
signals with a characteristic frequency fixed by the curvature
scale of the extra dimension---around $300\,$GHz or above---and
with amplitude comparable to the predictions of GR.  The most
efficient (and cost-effective) means of observing KK radiation
generated on our brane is with a high-frequency GW detector
optimised for all spins of radiation. The detection of such a
signal would provide a strong evidence in favour of new physics
(such as a large extra dimension), since there are no
astrophysical sources that can generate a similar spectrum of GWs.
(Certain models of inflation can generate GWs with $f \sim
10^{12}\,\text{Hz}$ \cite{Giovannini:1999bh}, but not with a
discrete spectrum.)

It is remarkable that both the intrinsic strength and shape of the
shadow matter GW spectrum implies that it is easier to detect than
KK radiation from visible matter.  In braneworld models, the only
means of directly observing material hidden in the bulk is via
gravitational interactions, so the search for KK radiation is one of
the few viable techniques that can constrain the shadow brane's
matter content. Intriguingly, our work suggests that real
observational constraints on shadow matter are within the grasp of
current technology such as LIGO.

We note that the discrete frequencies within the KK spectrum are
independent of the string and particle masses.  That is, all such
systems will generate GWs with the same frequencies, hence there
will likely be a significant (non-cosmological) integrated
stochastic GW background in this model.  Indeed, one would expect
such a background to be generated in any model incorporating
`ultraviolet' modifications to GR at length scales $\ell \lesssim
0.1\,$mm. Hence, stochastic GW backgrounds might prove to be a
useful model-independent observational window onto
submillimetre-scale exotic physics.

Finally, we reiterate that all of our results have been calculated
using the delta-function approximation for brane sources.  An
important open issue involves the effects of more sophisticated
source modeling, but this is left for future work.

\paragraph*{\bf Acknowledgements}

We would like to thank Chris Van Den Broeck and Roy Maartens for
discussions and comments, and Mike Cruise for insights into
high-frequency GW detectors. SSS is supported by PPARC.

\paragraph*{\bf Appendix}

In the Randall-Sundrum gauge \cite{Randall:1999ee}, perturbations of
the black string metric are orthogonal to the extra dimension and
given by
 $
    g_{\alpha\beta} \rightarrow g_{\alpha\beta} + \sum_{n=0}^\infty
    Z_n(y)h^{(n)}_{\alpha\beta}(x^\mu)
 $
. The linearized Einstein field equations yield that the $Z_n$ are
eigenfunctions of $-e^{-2|y|/\ell}(\di_y^2 - 4/\ell^2)$ with
discrete eigenvalues $m_n^2 c^2/\hbar^2$, which are the effective
masses of the KK gravitons $h_{\alpha\beta}^{(n)}$. The $n=0$
contribution is massless and hence reproduces ordinary GR. To avoid
the Gregory-Laflamme instability, we need $\mu_n = GMm_n/\hbar c >
0.4301$ for all $n >0$.

The individual components of the spherical part of
$h_{\alpha\beta}^{(n)}$ can be derived from a master variable
$\psi$, which satisfies
\begin{subequations}\label{wave equations}
\begin{eqnarray}
  \label{psi equation} ( \di_\tau^2 - \di_x^2 + V_\psi ) \psi & = &
  \mathcal{S}_\psi + \mathcal{{I}}\,\varphi, \\
  \label{phi equation} (\di_\tau^2 - \di_x^2 + V_\varphi) \varphi
  & = & \mathcal{S}_\varphi,
\end{eqnarray}
\end{subequations}
where we have defined the dimensionless coordinates $\tau=
{tc^3}/{GM}$ and $x = {rc^2}/{GM}+ 2 \ln\left( {rc^2}/{2GM}-1
\right)$. The tortoise coordinate $x$ maps the event horizon at $r =
2GM/c^2$ onto $x = -\infty$.  In these wave equations, $\varphi$ is
another master variable that governs spherical perturbations in the
position of the brane on which the matter source resides. The
equations are coupled by the interaction operator ${\mathcal{I}} =
{\mathcal{I}}(x,\di_x)$, and $V_{\psi,\phi}$ are potentials.

The source terms $\mathcal{S}_\psi$ and $\mathcal{S}_\varphi$
depend on the perturbing brane matter.  As in GR, we analytically
model a small brane particle using a stress-energy tensor with
delta-function support along its worldline.  In numeric
simulations, the delta-functions are replaced with a narrow
Gaussian profile~\cite{2003CQGra..20.3259L,2006CQGra..23..251S}.
There are a few ambiguities in this regularization scheme, but we
find that our numeric results far from the string are largely
insensitive to the particular choices made.

A detailed analysis leads to the following
late-time/distant-observer approximation for the KK metric
perturbations:
\begin{eqnarray}\nonumber
   \!\!\! h^{(n)}_{\alpha\beta} & \approx & h_n e^{i\omega_n t}
    \,\, \text{diag} \left( 0,+1,-\tfrac{1}{2}r^2,
    -\tfrac{1}{2}r^2\sin^2\theta \right)
    \\ && \times
    \begin{cases}
        1, & \text{periodic orbits},\\
        (tc^3/GM)^{-5/6}, & \text{fly-by orbits},
    \end{cases}
\end{eqnarray}
where
 $
    |h_n| = \sqrt{8\pi} \mathcal{A} \left( \frac{2GM_p}{rc^2}
    \right) \left(\frac{2GM}{\ell c^2}\right)^{-1/2} F_n(d/\ell)
 $
. $\mathcal{A}$ is a dimensionless quantity that depends on the
particle orbit but not on any other parameters; its value must be
determined from simulations. $F_n(d/\ell)$ is a complicated
expression with the following limiting behaviour:  When the
perturbing matter is on our brane
\begin{subequations}
\begin{equation}
    F_n(d/\ell) \approx
    \begin{cases}
        \tfrac{1}{2} e^{-3d/2\ell} (n\pi^3)^{1/2}, & n \ll 2 e^{d/\ell}/\pi^2, \\
        e^{-d/2\ell} (n\pi)^{-1/2}, & n \gg 2 e^{d/\ell}/\pi^2.
    \end{cases}
\end{equation}
On the other hand, for shadow particles
\begin{equation}
    F_n(d/\ell) \approx
    \begin{cases}
        e^{-d/2\ell} (\pi/2)^{1/2}, & n \ll 2 e^{d/\ell}/\pi^2, \\
        (n\pi)^{-1/2}, & n \gg 2 e^{d/\ell}/\pi^2.
    \end{cases}
\end{equation}
\end{subequations}
Finally, to a good approximation, the KK frequencies are given by
 $
    \omega_n = 2\pi f_n \approx \frac{c}{\ell} \left( n+\tfrac{1}{4} \right)
    \pi e^{-d/\ell}.
 $

\bibliography{bs}

\vfill

\end{document}